\documentstyle[sprocl,epsf]{article}

\input epsf
\begin{document}

\title{HYPERON POLARIZATION AND SINGLE SPIN ASYMMETRIES 
IN DIFFERENT HADRONIC REACTIONS 
\footnote{Supported in part by the National Natural Science 
Foundation of China (NSFC)}}
\author{LIANG ZUO-TANG}
\address {Department of physics, Shandong University, Jinan, Shandong 
250100, China}
\maketitle
\abstracts{ We show that the polarization of hyperons 
observed in high energy collisions using  
unpolarized hadron beams and unpolarized 
nucleon or nuclear targets 
is closely related to the left-right asymmetries 
observed in single spin inclusive hadron production processes.
The relationship is most obvious for the production 
of the hyperons which have only one common 
valence quark with the projectile. 
Further implications of the
existence of large polarization for hyperon 
which has two valence quarks in common with the projectile 
and their consequences are discussed.
A comparison with the available data is made. 
Further tests are suggested.}

This talk is a summary of the basic ideas and main results 
of Refs.[1-3] which have been done 
in collaboration with C.~Boros.
In these papers, we show that there should be 
a close relation between 
hyperon polarization $(P_H)$ and single-spin 
left-right asymmetry $(A_N)$ 
observed in hadron-hadron 
collisions at high energies.
We pointed out that following simple observations 
and/or arguments seem already suggest this should be true:
First, the data [4] for $P_H$ and those for $A_N$  
show exactly the same characteristics. 
Second, both effects show the existence of 
a correlation between transverse motion 
and transverse polarization.  
Hence, unless the polarization of 
the produced hyperons in 
the projectile fragmentation region  
(denoted by PFR in the following)
is completely independent of that of the projectile 
we are practically forced to accept that 
$A_N$ and $P_H$ are closely related to each other.  

For explicity, we considered the questions: 
Can we understand the 
existence of $P_H$ and reproduce the main feature of the data if we use 
$A_N$ as input? Do we need further input(s)?
We noted in particular the following:

(I) The $A_N$ data show that meson containing $q_v$ 
and a suitable anti-sea-quark $\bar q_s$ 
has a large probability to go left 
if $q_v$ is upwards polarized.
(We denote the difference of 
the probability for such a meson to go left 
and that to go right by $C$.] 
We assume that this is also true for the produced baryon 
which contains such a $q_v$ and a sea diquark. 

(II) Recent measurements [5] 
of $\Lambda$ polarization from $Z^0$ decay show that,
in the longitudinally polarized case,  
quark polarization remains the same 
before and after the hadronization. 
We assume that this is true 
not only for $\Lambda$  
but also for other hyperons 
and also in the transversely polarized case. 

We first considered the production of 
hyperon $(H)$ which has only 
one valence quark in common with the projectile $P$. 
In this case, the PFR is dominated 
by the hadronization product that contains 
this common valence quark,  and the polarization 
of such $H$ is determined completely by the 
points (I) and (II) mentioned above.  
E.g., $p+p\to \Sigma^-+X$, 
the common valence quark is $d_v$ 
and, according to (I), 
$d_v$ should have a large probability 
to be upwards polarized, if $\Sigma^-$ is going left. 
Together with the wavefunction of $\Sigma^-$, 
this implies that the produced $\Sigma^-$ 
has a probability of 
$5/6 (1/6)C$ to be upwards (downwards) polarized, 
i.e. $P_H=(2/3)C$ for such $\Sigma^-$.
Similarly, we obtain $P_H=(-1/3)C$ 
for such $\Xi^-$ or $\Xi^0$ 
in $pp$-collisions; and so on. 
Since such $H$'s dominate only at large $x_F$,
we expect that the magnitudes of $P_H$ 
increase with increasing $x_F$ and the 
above mentioned results are their 
limits at $x_F\to 1$. 
All these are consistent with the data [4].
Furthermore, without any other input, 
we obtained, e.g.:
(A) $P_\Lambda$ in the PFR  
    of $K^-+p\to \Lambda+X$ 
    is large and is, in contrast to 
    that in $pp$-collisions, positive in sign.
(B) $P_\Lambda$ in the PFR of 
  $\pi^\pm+p\to \Lambda +X$ should be negative 
  and the magnitude should be very small.
(C) Not only hyperons 
 but also the produced vector mesons 
 are expected to be transversely polarized
 in the fragmentation region 
 of hadron-hadron collisions.
(D) Neither the contribution from 
hadronization to $P_\Lambda$ nor 
that to $A_N$ can be large.  
Presently, there are already data available 
for (A) and (B) [4], 
and they are {\it in agreement with 
these associations}. 
(D) is consistent with the results [5] of the recent 
SLD measurements of jet handedness at SLAC, which show 
that the spin dependence of hadronization is very little.
(C) can be checked by future experiments.

In the second case, we considered the 
production of hyperon which has two valence quarks 
in common with the projectile and hence
hyperons containing such common valence diquarks 
dominate the PFR. 
E.g., $p+p\to \Lambda +X$. 
We started [2] again from $A_N$ for 
$p(\uparrow)+p\to \Lambda+X$ and recall
that a significant $A_N$ has been observed [4]
in the PFR.  
This result was rather surprising since, 
according to the wavefunction of $\Lambda$, 
the common valence diquark $(u_vd_v)$ here 
has to be in the spin zero state thus cannot 
transfer the information of 
polarization to the produced $\Lambda$.
An explanation is given in [2].
It has been pointed out [2] that, 
the production of the $\Lambda$ containing 
the spin-0 $(u_vd_v)$ and a $s_s$ is
associated with the production of a $K$ 
containing the remaining $u_v^a$ of $P$
and the $\bar s_s$. 
The information of polarization
is carried by the $u_v^a$
so that, according to (I), 
the produced $K$ has a large probability to 
go left if $P$ is upwards polarized. 
The $\Lambda$ has therefore 
a large probability to go right 
since the transverse momentum should be compensated. 
In this way, the $A_N$ data [4] has been reproduced. 
According to this,  in unpolarized $pp$-collision, 
if the produced $\Lambda$ is moving to the left,  
the associated $K$ should mainly move to the right. 
Hence, according to (I), 
the $u_v^a$ contained in this $K$ should have 
a large probability to be downwards polarized. 
Since $K$ has spin 0, the $\bar s_s$ should
be upwards polarized. 
Hence, to get a negative $P_\Lambda$, 
we need only to assume 
that the sea quark-antiquark pair $s_s\bar s_s$ 
have opposite transverse spins.  
Under this assumption, the polarization of such 
$\Lambda $ is completely determined by that 
of the remaining $u_v^a$. 
To test this, we did:
First, we made a
similar analysis for other hyperons. 
The obtained $P_H$'s   
are all consistent with the available data~[4]. 
Second, we made a quantitative estimation 
of $P_\Lambda$ in $p+p\to\Lambda+X$ 
as a function of $x_F$. 
By taking the $\Lambda$'s 
containing two, one or zero valence quark(s) of $P$
into account, we obtained the result shown in Fig.1.  
Third, we derived a number of other consequences without any further input. 
E.g.:
($\alpha$) The spin transfer 
$D_{NN}$ in the PFR of
$p(\uparrow)+p\to \Lambda+X$ 
should be positive and large for large $x_F$.   
($\beta$) $P_\Lambda$ in the PFR
of $\Sigma^-+A\to \Lambda +X$ 
should be {\it negative} and much less significant 
than that in $p+p\to\Lambda+X$.  
($\gamma$) Hyperon polarization in  
processes in which a vector meson is associatively produced 
should be very much different from that in  
processes in which a pseudoscalar 
meson is associatively produced.  
Presently, there are data available 
for the processes mentioned 
in ($\alpha$) and ($\beta$) [4], 
and they are in agreement with 
the above expectations. 
($\gamma$) is another characteristic feature of the model 
and can be used as a crisp test of the picture.

\begin{tabular}{ll}
\begin{minipage}[t]{6cm}
\epsfxsize=6truecm \epsfbox{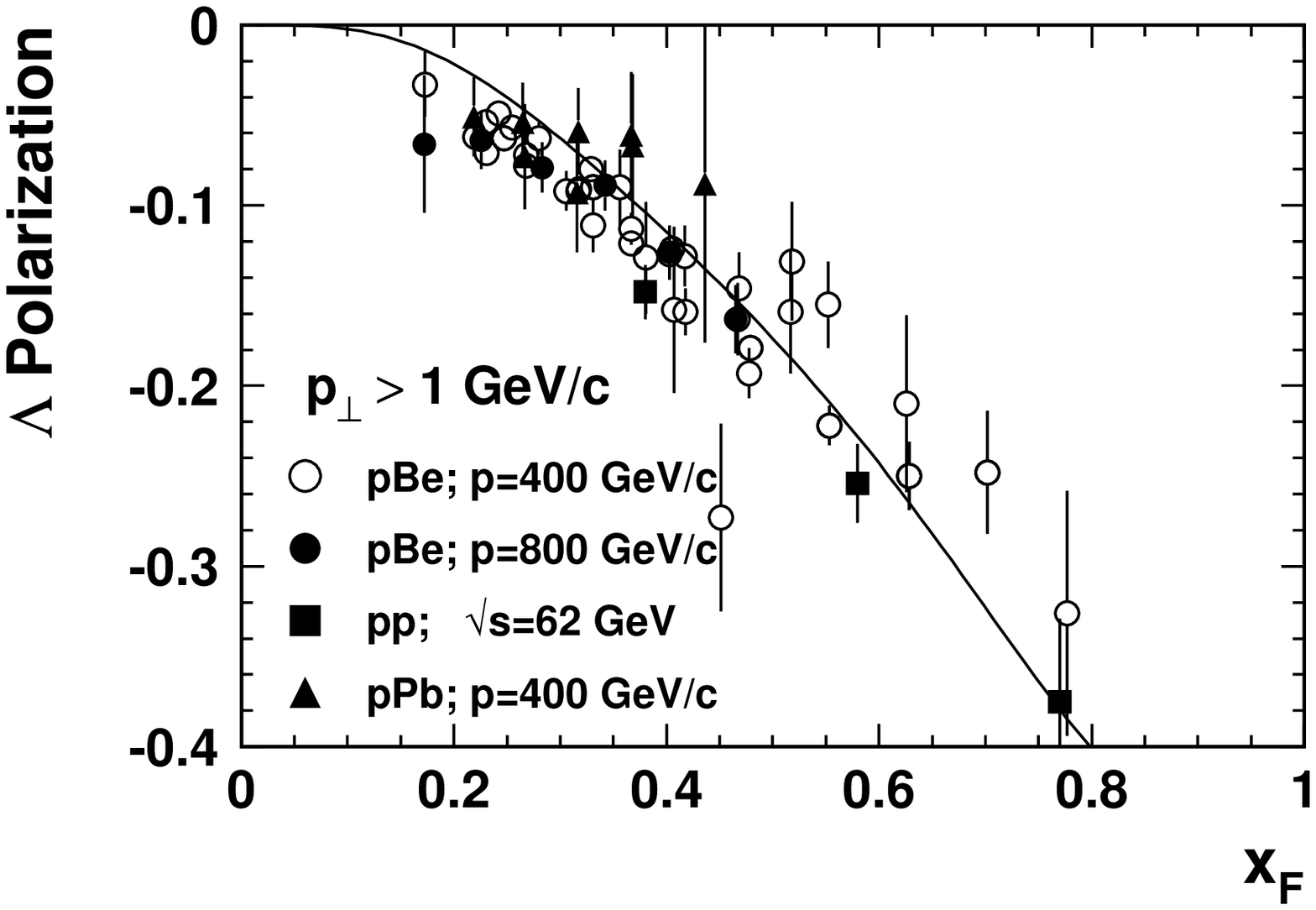}
\end{minipage}
&
\begin{minipage}[b]{4cm}
Fig.1: Calculated results for the polarization of $\Lambda$, 
$P_\Lambda$, as a function of $x_F$ compared with the data [4].
See Ref.[1] for more detail.

\vspace{1cm}
\end{minipage}
\\
\end{tabular}

\end{document}